\documentclass[twocolumn]{emulateapj}
\usepackage[dvips]{epsfig} 
\usepackage[dvips]{rotating}
\usepackage{subfigure}
\renewcommand{\thefootnote}{\fnsymbol{footnote}}

\shorttitle{The Metallicity of the Monoceros Stream}
 
\shortauthors{Meisner et al.} 
 
\begin{document} 
\title{The Metallicity of the Monoceros Stream\footnote{O\MakeLowercase{bservations reported here were obtained at the }MMT O\MakeLowercase{bservatory, a joint facility of the }S\MakeLowercase{mithsonian }I\MakeLowercase{nstitution and the }U\MakeLowercase{niversity of }A\MakeLowercase{rizona.}}}

\author{Aaron M. Meisner\altaffilmark{1,2}}
\author{Anna Frebel\altaffilmark{2,3}}
\author{Mario Juri{\'c}\altaffilmark{2,4}}
\author{Douglas P. Finkbeiner\altaffilmark{1,2}}
\altaffiltext{1}{Department of Physics, Harvard University, 17 Oxford Street, Cambridge, MA 02138, USA; ameisner@fas.harvard.edu}
\altaffiltext{2}{Harvard-Smithsonian Center for Astrophysics, 60 Garden St, Cambridge, MA 02138, USA; mjuric@cfa.harvard.edu, dfinkbeiner@cfa.harvard.edu}
\altaffiltext{3}{Present address: Massachusetts Institute of Technology, Kavli Institute for Astrophysics and Space Research, 77 Massachusetts Avenue, Cambridge, MA 02139, USA; afrebel@mit.edu}
\altaffiltext{4}{Hubble Fellow}

%abstract is on the "title page"
\begin{abstract} 
We present low-resolution MMT Hectospec spectroscopy of 594 candidate
Monoceros stream member stars. Based on strong color-magnitude diagram
overdensities, we targeted three fields within the stream's footprint,
with $178^{\circ}\leq l \leq 203^{\circ}$ and $-25^{\circ} \leq b \leq
25^{\circ}$. By comparing the measured iron abundances with those
expected from smooth Galactic components alone, we measure, for the
first time, the spectroscopic metallicity distribution function for
Monoceros. We find the stream to be chemically distinct from both the
thick disk and halo, with $\mbox{[Fe/H]}=-1$, and do not detect a trend in
the stream's metallicity with Galactic longitude. Passing from
$b$=$+25^{\circ}$ to $b$=$-25^{\circ}$, the median Monoceros
metallicity trends upward by 0.1 dex, though uncertainties in modeling
sample contamination by the disk and halo make this a marginal
detection. In each field, we find Monoceros to have an intrinsic
[Fe/H] dispersion of $0.10$-$0.22$ dex. From the Ca\,II\,K line, we
measure [Ca/Fe] for a subsample of metal poor program stars with
$-1.1<\mbox{[Fe/H]}<-0.5$. In two of three fields, we find calcium
deficiencies qualitatively similar to previously reported [Ti/Fe]
underabundances in Monoceros and the Sagittarius tidal
stream. Further, using 90 spectra of thick disk stars in the Monoceros
pointings with $b \approx \pm25^{\circ}$, we detect a 0.22 dex
north/south metallicity asymmetry coincident with known stellar
density asymmetry at $R_{\rm GC} \approx 12$ kpc
and $|Z| \approx 1.7$ kpc. Our median Monoceros $\mbox{[Fe/H]} =-1.0$ and its
relatively low dispersion naturally fit the expectation for an
appropriately luminous $M_V \sim -13$ dwarf galaxy progenitor.
\end{abstract}  
 
 
\section{Introduction} 
 
The Monoceros stream \citep[][hereafter Y03]{newberg02,yanny03}
comprises a ring-like stellar overdensity in the plane of the Galactic
disk. Between $110^{\circ} \lesssim l \lesssim 250^{\circ}$, the
kinematically cold (Y03) stream covers $-30^{\circ} \lesssim b
\lesssim 30^{\circ}$ at galactocentric radius R$\sim$17-19 kpc with
radial thickness $\Delta R \sim 4$ kpc \citep{juric08}.

The true nature of the structure is still uncertain, with the leading
scenarios being: a) remnant of an accretion event \citep{penarrubia},
with the Canis Major dwarf galaxy \citep{martin} as a possible
progenitor, b) disturbance due to a high-eccentricity flyby encounter
\citep{younger}, and c) disk flare or warp \citep[e.g.,][]{momany}

Discerning between these scenarios, especially the first two, may have
important theoretical consequences as mergers with orbits and mass
ratios implied by \cite{penarrubia} are deemed unlikely in current
$\Lambda$CDM models \citep[e.g.,][]{younger}.

In characterizing any stellar population, metallicity is considered a
fundamental parameter. Unfortunately, the literature regarding the
chemical composition of the Monoceros stream is sparse and seemingly
inconsistent. For example, Y03 estimate $\mbox{[Fe/H]} \sim -1.6$ at $l =
198^{\circ}$, $b = −27^{\circ}$ using the Ca\,II\,K line
and colors of turnoff stars. On the other end, \cite{crane}
(henceforth C03) measure a mean of $-0.4\pm0.3$ using M giants,
sampled mainly from observations of the northern component of the
stream. Most recently, using photometric metallicities, \cite{i08}
(hereafter I08) find the MSTO stars in the northern anticenter portion
of the stream to have a median metallicity of $\mbox{[Fe/H]} = -0.95$, with
intrinsic RMS scatter of only 0.15 dex. Hypotheses invoked to
reconcile these values range from multiple stellar populations
\citep[e.g.,][]{penarrubia}, to systematic errors in calibrations
(C03), and to questioning whether the northern and southern part of
the ring share the same origin \citep{conn07}. Also, except for the
I08 photometric study, the full metallicity distribution function
(MDF) of stars in the Monoceros stream has actually never been
measured from spectroscopy.

To better characterize the chemical composition of Monoceros, we have
obtained spectra of $\sim$$600$ candidate member stars. This large
sample size allows us to probe the Monoceros MDF with good
statistics. Also, by pointing in multiple directions along the stream,
we can search for metallicity gradients. Specifically, we set out to
characterize the [Fe/H] MDF of Monoceros in each of three fields with
$177.9^{\circ}\leq l \leq 202.9^{\circ}$ and $-24.2^{\circ} \leq b
\leq 24.4^{\circ}$ (see Table \ref{obs} for field details and naming
conventions). Additionally, we aim to gauge the $\alpha$-enhancement
of Monoceros from a [Ca/Fe] measurement. This allows us to address the
origin of Monoceros by comparing its [Ca/Fe] to the differing
$\alpha$-abundance patterns found in the Milky Way versus dwarf
spheroidal (dSph) galaxies.

\begin{deluxetable*}{rrrrrrrrrrr} 
\tabletypesize{\scriptsize}
\tablecolumns{10} 
\tablewidth{0pc} 
\tablecaption{\label{obs} Summary of MMT Observations/Targets} 
\tablehead{
\colhead{Sample} &
\colhead{$l$} & 
\colhead{$b$} & 
\colhead{$\alpha$} & 
\colhead{$\delta$} &
\colhead{Exposures (s)} &
\colhead{N$_{stars}$} & 
\colhead{(S/N)$_{med}$} & 
\colhead{$r_{med}$} & 
\colhead{$|Z|_{med}$\tablenotemark{a} (kpc)} &
\colhead{$A_r$}
}
\startdata
\multicolumn{11}{c}{NORTH} \\
Monoceros & 187.245$^{\circ}$ & 24.401$^{\circ}$ & 115.741$^{\circ}$ & 32.662$^{\circ}$ & 2$\times$7200 & 199 & 49 & 19.55 & 3.871 & 0.12 \\
Thick Disk & 187.245$^{\circ}$ & 24.401$^{\circ}$ & 115.741$^{\circ}$ & 32.662$^{\circ}$ & 2$\times$7200 & 30 & 123 & 17.62 & 1.603 & 0.12 \\
\multicolumn{11}{c}{SOUTH} \\
Monoceros & 202.884$^{\circ}$ & $-$24.183$^{\circ}$ & 76.908$^{\circ}$ & $-$2.587$^{\circ}$ & 2$\times$7200 & 146 & 47 & 19.83 & 4.911 & 0.30 \\
Thick Disk & 202.884$^{\circ}$ & $-$24.183$^{\circ}$ & 76.908$^{\circ}$ & $-$2.587$^{\circ}$ & 2$\times$7200 & 60 & 137 & 17.44 & 1.788 & 0.30 \\
\multicolumn{11}{c}{NORTH18} \\
Monoceros & 177.879$^{\circ}$ & 18.055$^{\circ}$ & 104.834$^{\circ}$ & 38.852$^{\circ}$ & 2$\times$7200,5400 & 249 & 84 & 19.08 & 2.688 & 0.29 \\
\multicolumn{11}{c}{M13} \\
Calibration & 58.999$^{\circ}$ & 40.917$^{\circ}$ & 250.416$^{\circ}$ & 36.454$^{\circ}$ & 3600 & 189 & 39 & 18.34 & 5.028 & 0.05 \\ 
\enddata
\tablenotetext{a}{using the photometric parallax relation of I08 equations (A6) and (A7)} 
\end{deluxetable*}

\section{MMT Observations}\label{sec:obs} 
\subsection{Sample Selection}

Recognized as a stellar overdensity between $b=\pm 30^{\circ}$ for
$110^{\circ} \lesssim l \lesssim 250^{\circ}$, Monoceros appears most
strikingly in the number counts of MSTO F/G dwarfs at galactocentric
distances R$\sim$15-20 kpc. Near the Galactic anticenter, these stars
correspond to apparent magnitude $r \approx 18.5$-$20.5$ and color
$g-r \approx 0.25$-$0.4$, with $ugriz$ referring to SDSS photometric
passbands. By examining ($g$, $g-r$) color-magnitude diagrams (CMDs)
of objects classified by the SDSS (DR7) as stars within the joint
Monoceros/SDSS footprint, we identified many degree-scale pointings in
which the Monoceros MSTO was obviously visible above the Galactic
background (for background estimation details see
$\S$\ref{sec:decomp}). In the best cases, the relevant CMD region of
such pointings exhibited a factor of $\sim$$2$ overdensity, with total
projected density $\sim$$200$$-$$300$ stars/deg$^2$. This field size
and target density matches that of MMT/Hectospec, which, with its
large 6.5m mirror size, can obtain at $r=20$ S/N sufficient to measure
[Fe/H] with integration times of $\sim$4 hours.

Hoping to gauge both $l$ and north/south gradients in the stream's
metallicity, six such fields were initially targeted for MMT/Hectospec
spectroscopy, covering $150^{\circ} \lesssim l \lesssim 230^{\circ}$
in the Galactic north and sampling four Galactic latitudes with
$-30^{\circ} < b < 30^{\circ}$. In several of these fields, a
relatively small subsample of much brighter thick disk targets were
chosen from a similar range of MSTO $g-r$ colors and with $17<r<18$.

\subsection{Observations and Data Reduction} 

\begin{figure*}[ht]
\centering
\subfigure[]{
\includegraphics[scale=0.54]{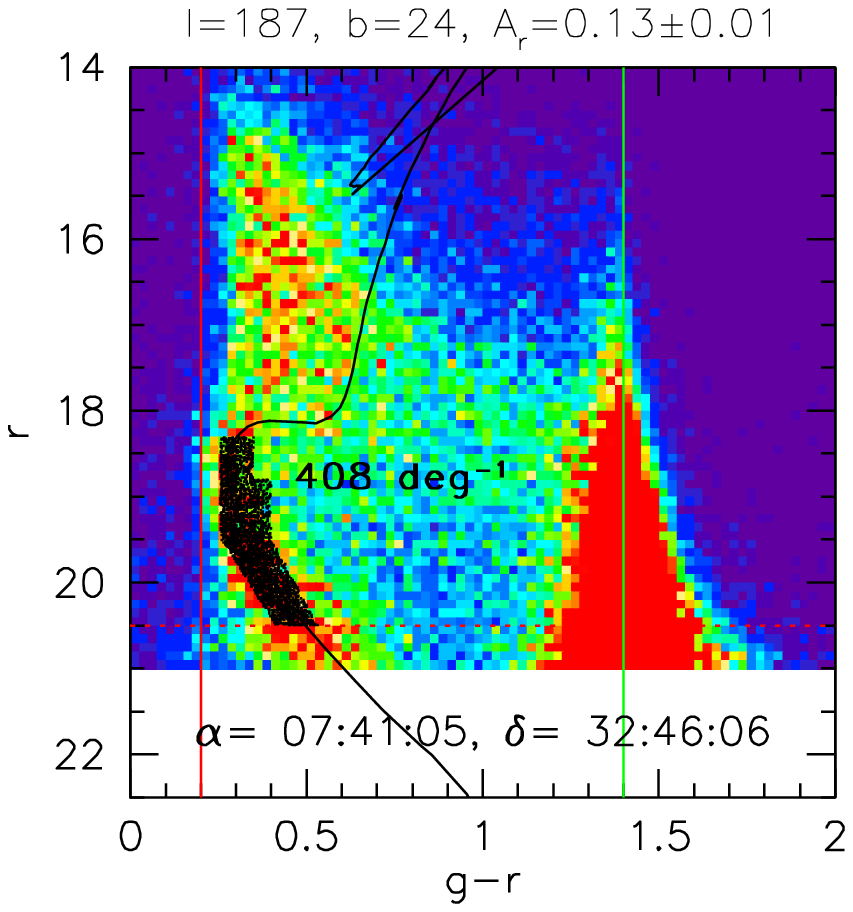}
}
\subfigure[]{
\includegraphics[scale=0.54]{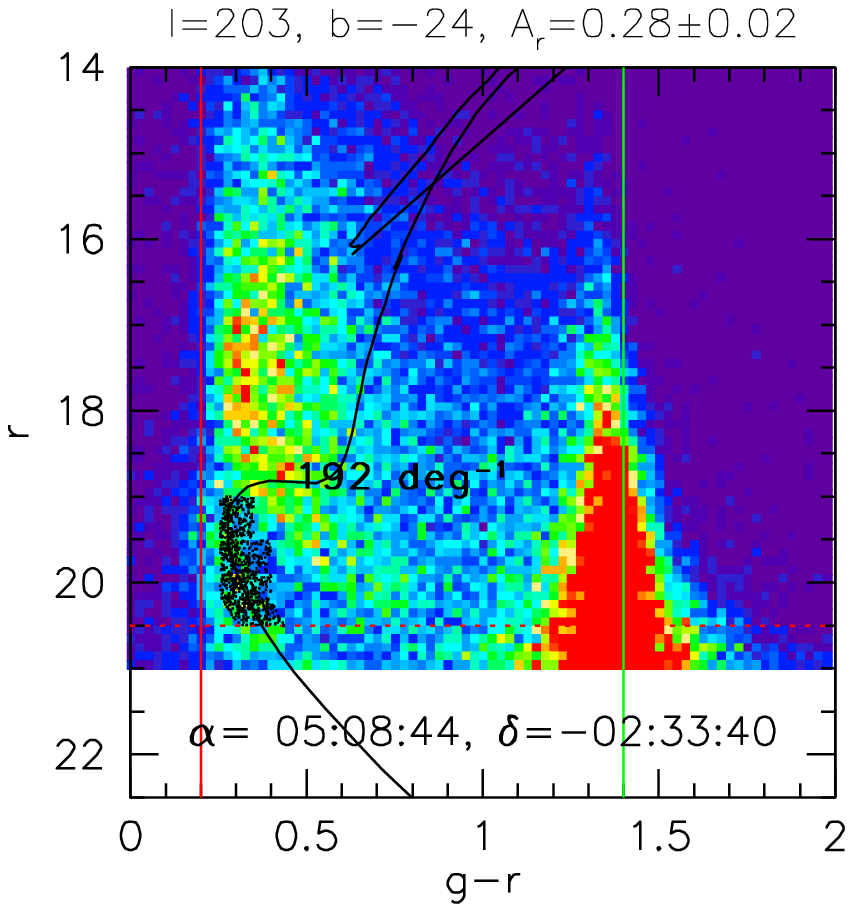}
}
\subfigure[]{
\includegraphics[scale=0.54]{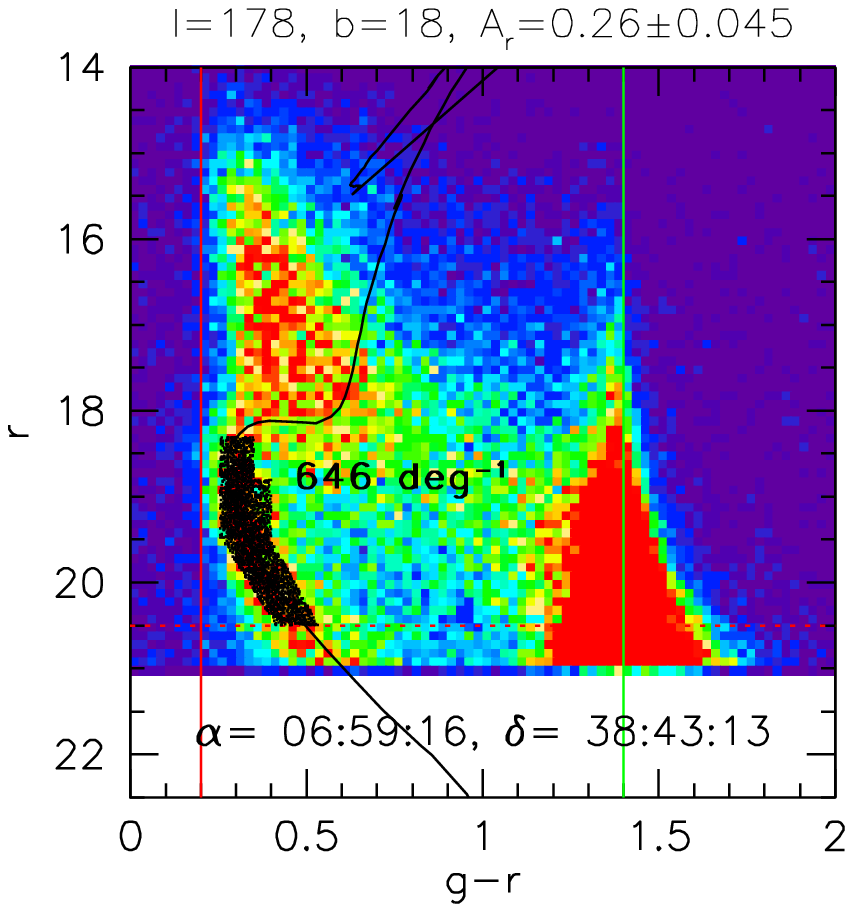}
}
\caption[]{\label{cmd} SDSS DR7 CMDs of target fields NORTH (a), SOUTH
(b) and NORTH18 (c). All three panels have the same dynamic range
(0-30 stars/pixel, with the pixel sizes being 0.025 mag and 0.1 mag in
$g-r$ and $r$, respectively). The black dots represent Monoceros
candidates from which the spectroscopic sample we analyzed was
drawn. The black line shows an 8 Gyr, $Z=0.0039$ isochrone
\citep{marigo} intended only to guide the eye along the Monoceros
MSTO.}
\end{figure*}

The observations were carried out in MMT queue-scheduled multi-object
observing mode. Due to a combination of factors including weather,
only three fields were observed, spanning a limited range of ($l$,
$b$), though still including both Galactic north and south. Table
\ref{obs} summarizes the targets/observations, and Figure \ref{cmd}
shows the CMD of each field, indicating targeted regions.

Hectospec can observe up to 300 objects at once. In general, we
typically devoted $\sim$50 fibers to measuring the sky, with the
remainder targeting Monoceros candidates subject to the constraints of
the instrument (e.g., the availability of the targets in the desired
magnitude range, and limits on how closely spaced the fibers could
get). Any remaining fibers were set to target the thick disk stars, to
investigate the thick disk asymmetry present towards the Galactic
anti-center (see $\S$\ref{thickdisk}). We used the 270 gpm grating
yielding a resolving power of $R\sim1$,000 near $5000
\textrm{\AA}$. The low resolution warrants full wavelength coverage of
$3700$-$9150\textrm{\AA}$.
 
The data were reduced with customized IRAF procedures as part of the
SAO Telescope Data Center service \citep{mink07}.  The S/N ratio of
the final spectra is typically $\sim$50-80 per pixel at
$\sim5200\textrm{\AA}$ for the Monoceros stars and $\sim$135 for the
brighter thick disk stars. We used the IRAF/rvsao routine \verb|xcsao|
to measure the radial velocity ($v_{rad}$) of each star from
cross-correlation with template stellar spectra on a per-exposure
basis, correcting for heliocentric velocity before stacking the
exposures of each object. Typical errors on $v_{rad}$ ranged from
10-20 km~s$^{-1}$, varying based on S/N.

\subsection{Stellar Parameters} 

Abundance extraction requires estimates of several stellar parameters
of each object, namely: the surface gravity log$g$, temperature
$T_{\rm eff}$, microturbulence $v_t$, and
metallicity. As our targets were selected from SDSS imaging, each star
has $ugriz$ photometry available. Using I08 equation (3), which gives
$T_{\rm eff}$ as a function of $g-r$, we
determined effective temperatures for all program stars. The
statistical error per star due to photometric errors is $\sim$$80$K
for thick disk targets and $\sim$$110$K per star for the fainter
Monoceros candidates. $T_{\rm eff}$ generally
ranged from 5800-6300K within each sample of Monoceros/thick disk
stars.

Our targets were selected to be MSTO F/G dwarfs. Based on the stellar
parameters of 46 dwarfs from \cite{fulbright} with
$-2 \le \mbox{[Fe/H]} \le 0.0$ and $5800$ K$\le T_{\rm eff}
\le$6300 K, we adopt $v_t = 1.25$ km~s$^{-1}$ and log$g=4.20$ for all
Monoceros candidates and thick disk stars. In $\S$\ref{sec:err} we
discuss the impact of these assumptions.

Of course, we wish to derive rather than assume metallicities. To
extract abundances, we first determine [Fe/H] from iron absorption
features assuming [X/Fe]=0 for all X, choosing features where this
assumption has negligible effect (see Figure \ref{fe_plot} and
$\S$\ref{sec:err}). The [Fe/H] value of each object is then used as
the metallicity input for subsequent Ca abundance measurements.

\section{Iron Abundance of the Monoceros Stream}\label{ab_fe} 

\begin{figure*}[ht]
\centering
\subfigure[]{
\includegraphics[scale=0.3875]{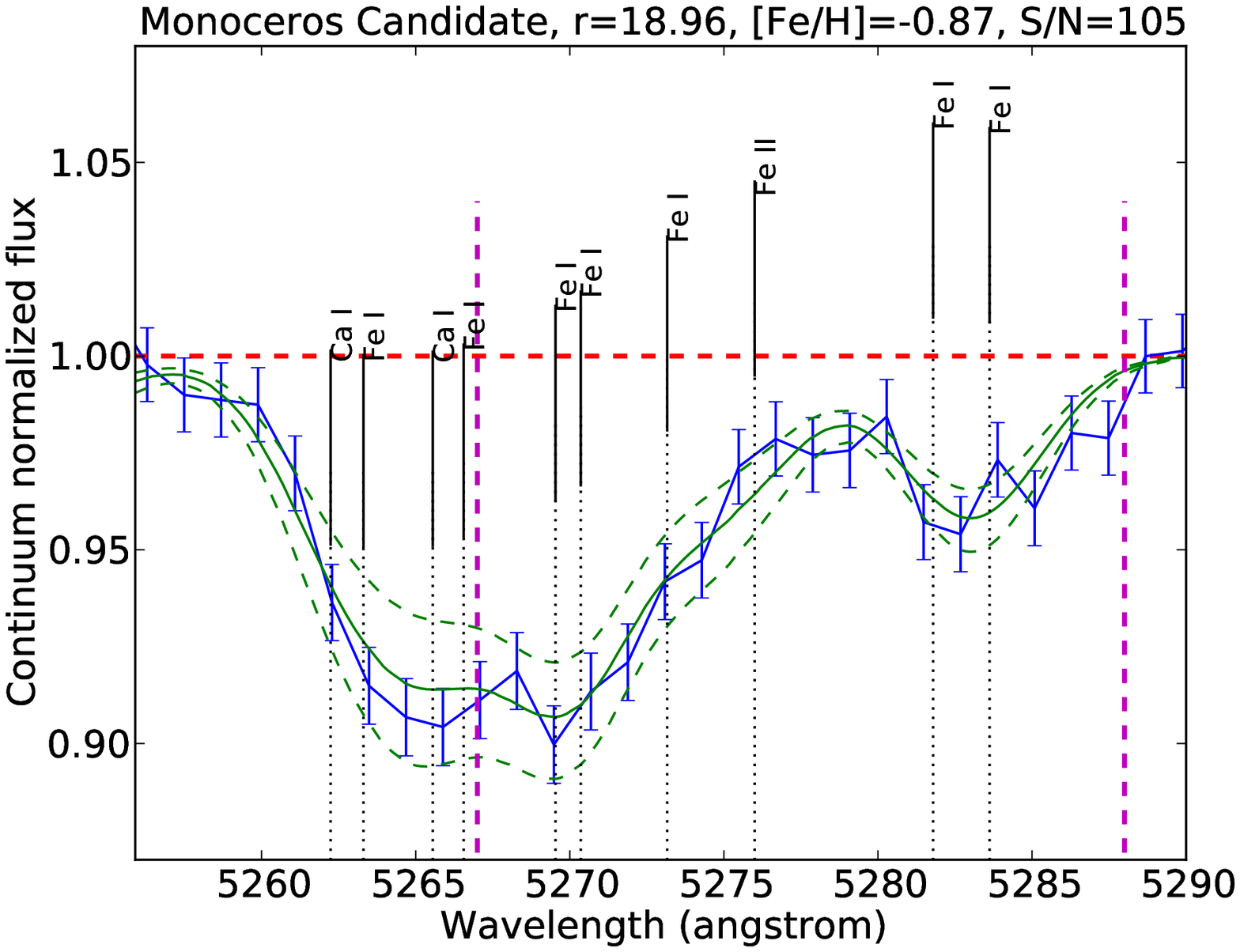}
\label{fe_plot}
}
\subfigure[]{
\includegraphics[scale=0.3875]{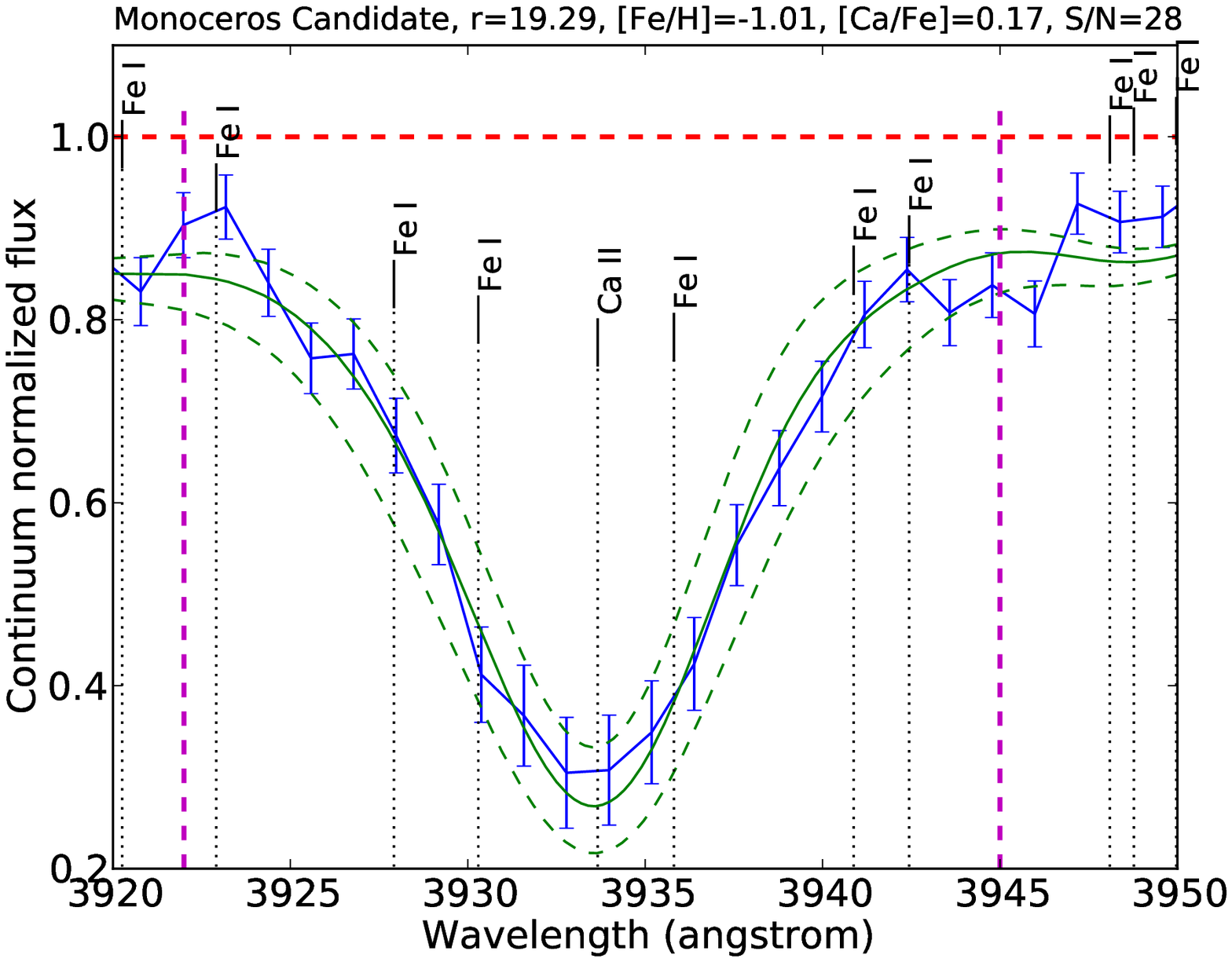}
\label{ca_plot}
}
\caption[]{(a) An example of the [Fe/H] fitting region for typical
[Fe/H]. The blue line with errorbars is the observed data, the green
line is the best-fit synthetic spectrum, the red dashed line is the
continuum fit, and the purple dashed lines show the fitting region
boundary. The upper/lower green dashed lines show synthetic spectra
with [Fe/H] lowered/raised by 0.25 dex relative to the best fit. The
strong Fe\,I line at 5269.5$\textrm{\AA}$, plus the next 9 most
important lines by EW are marked with black vertical lines. (b) Same
for Ca\,II\,K at 3933.7$\textrm{\AA}$, plus next ten most important
lines by EW, all of which are Fe\,I features. The upper/lower green
dashed lines correspond to offsets in [Ca/Fe] of $\mp$0.25 dex
relative to the best fitting [Ca/Fe], keeping [Fe/H] fixed.}
\end{figure*}

To infer iron abundances from our spectra, we focus on the iron line
complex near $\lambda \approx 5270 \textrm{\AA}$. The absorption
equivalent width (EW) in the region with $5267 \textrm{\AA} < \lambda
< 5287\textrm{\AA}$ is dominated by Fe, mainly Fe\,I, but with some
Fe\,II. The primary Fe lines of interest and major contaminating lines
due to other elements are labeled in Figure \ref{fe_plot}. All
abundances, including those drawn from the literature for calibration,
were standardized with respect to the current solar values of
\cite{asplund}.

\subsection{Continuum Normalization}

Given our relatively low resolution and moderate S/N, continuum
normalization is a sensitive matter. We base our continuum fitting
procedure on that used by the SEGUE Stellar Parameter Pipeline
\citep[SSPP,][]{lee08} to measure [Fe/H], since our spectra have
similar S/N and resolution to those for which their pipeline was
designed. In particular, we iteratively fit the region of the spectrum
blueward of $5800\textrm{\AA}$ with a ninth order polynomial, masking
the strong Balmer lines. The SSPP additionally computes abundances
with local continuum estimates based on sidebands. However, our
resolution is a factor of $\sim$2 lower than that of SSPP spectra,
making it unfeasible to isolate sideband regions which attain the
continuum and are sufficiently wide to be robust.

\subsection{[Fe/H] Fitting Technique}\label{fe_fit}

We take the calculated $T_{\rm eff}$ and assumed
log$g$, $v_t$ for each star and compare the observed spectrum to a
library of synthetic spectra generated based on these fixed stellar
parameters, but with varying [Fe/H]. In particular, synthetic,
continuum normalized spectra were generated using MOOG \citep[the
latest version, 2010;][]{MOOG}, with one-dimensional no-overshoot
model atmospheres (assuming local thermodynamic equilibrium) from
\cite{kurucz93}.  This version of MOOG accounts for the fact that
Rayleigh scattering becomes an important source of continuum opacity
at short wavelengths blueward of 4500\,{\AA} \citep{sobeck11}. For
each [Fe/H] value with $-3.5 < \mbox{[Fe/H]} <0.5$ in gradations of 0.01 dex,
we compute the $\chi^2$ between the synthetic spectrum and the
observed, normalized spectrum over the Fe absorption region. We assign
each star the [Fe/H] corresponding to the minimum $\chi^2$ value thus
obtained.  The $\chi^2$ is computed with the variance taken from
Poisson statistics to be $\sigma^2=(N/g+R^2)$, where $N$ is the counts
per pixel in ADU, $R=2.8$ ADU is the readnoise and $g=1$ is the
gain. We compute the $\chi^2$ within the range $5267 \textrm{\AA} <
\lambda < 5287\textrm{\AA}$, a region determined to optimize the Fe
absorption relative to absorption by other elements and random
noise. An example best-fit result for typical S/N and iron absorption
strength is plotted in Figure \ref{fe_plot}. The line list employed in
our syntheses is provided in Table \ref{linelist}.

\subsection{Iron Abundance Calibration}

To test our [Fe/H] fitting procedure, we apply the above methodology
to 189 MMT/Hectospec spectra of globular cluster stars in M13 with
comparable S/N and SDSS photometry. We again estimate
$T_{\rm eff}$ from $g-r$ and adopt $v_t=1.5$
km~s$^{-1}$, log$g$ = 4.25 based on \cite{briley01} Table 2. M13 is
known to have $\mbox{[Fe/H]}=-1.55$ \citep{cohen}, and we recover a median
[Fe/H] of $-1.51$. We thus apply a $-0.04$ dex correction factor to
all of our [Fe/H] measurements.

\subsection{Rejecting Poorly Determined [Fe/H]}

Based on visual inspection of the faintest and apparently most metal
poor objects in our Monoceros candidate sample, it became clear that
several of these lacked sufficient iron EW to measure [Fe/H] in the
presence of statistical noise and minor continuum placement errors. To
remove these outliers, we computed $\sigma_{EW}$, the $1\sigma$ error
in total Fe fitting region EW due to Poisson noise and continuum
placement errors. The continuum error was estimated by fitting a
polynomial two orders lower than the default continuum, and generally
amounted to $<$$1\%$ continuum level difference. We then rejected all
stars for which the S/N and EW were sufficiently low that
$\sigma_{EW}$ translated to $>$0.5 dex [Fe/H] uncertainty. For
Monoceros targets in fields NORTH, SOUTH, NORTH18 the percentage of
stars rejected was 7.5\%, 6.8\% and 0.4\% respectively. All rejected
stars had estimated $\mbox{[Fe/H]} < -2.05$. No thick disk [Fe/H] estimates
were rejected, as these stars generally have higher S/N and
metallicity than the Monoceros candidates.

In performing maximum likelihood fits to our [Fe/H] distributions (see
$\S$\ref{sec:decomp}) we also identified, among Monoceros candidates,
three [Fe/H] outliers at $\mbox{[Fe/H]} > 0$. These objects also displayed
unreasonably low $\mbox{[Ca/Fe]} < -1$, again suggesting a dramatic
overestimate of [Fe/H]; these three objects were also excluded from
the sample.

\subsection{Iron Abundance Uncertainties}\label{sec:err}
There are several uncertainties on our derived metallicities. First
there is statistical scatter from Poisson noise, which, for typical
S/N is a function of [Fe/H]. We gauge this error with Monte Carlo
generation of spectra to which we have added random Poisson noise. For
typical $\mbox{[Fe/H]} = -1$, we find an RMS scatter of $0.11$ dex. At typical
halo $\mbox{[Fe/H]} = -1.6$ we find a scatter of 0.25 dex, while for typical
thick disk $\mbox{[Fe/H]}=-0.7$ we find an RMS scatter of only 0.05 dex. With
sample sizes of 100$-$250 stars per field, this statistical error is
thus not a primary concern.

Another source of uncertainty stems from our assumptions about stellar
parameters. We find the effects of reasonably large systematic
over/underestimates in each of log$g$ ($\pm 0.3$ dex),
$T_{\rm eff}$ ($\pm 100$ K), and $v_t$ ($\pm 0.3$
dex), to be $\mp 0.05$ dex, $\pm 0.07$ dex, $\pm 0.07$ dex in [Fe/H]
respectively. Summing these effects in quadrature yields an
uncertainty of $\pm 0.11$ dex on our derived metallicities.

A further possible source of bias in [Fe/H] is contamination of the
iron abundance fitting region by lines of other elements, most notably
Ca (see Figure \ref{fe_plot}). In determining [Fe/H], we generated
synthetic spectra assuming that [X/Fe]=0 for all X. While [X/Fe] is
not known for Monoceros, the most relevant parameter of [Ca/Fe] for
the Galactic disk/halo tends to lie between 0.0 and 0.4. We address
the resulting [Fe/H] bias by using MOOG to simulate populations of
stars with [Ca/Fe] following the known Galactic trend with [Fe/H] (see
$\S$\ref{ca_cont}), adding appropriate noise to the synthetic spectra
before feeding them back into our [Fe/H] fitter. The resulting bias is
very small, $\lesssim$0.02 dex.

A final important source of uncertainty is the continuum normalization
procedure. With our low resolution, the blending of lines means the
continuum is rarely achieved, and could potentially be systematically
over/underestimated throughout the sample, thereby raising/lowering
all metallicities. To test the systematic effect due to differing
continuum normalizations, we again re-fit abundances with new continua
parameterized by a polynomial two orders lower than the default
continuum. We find in general a resulting systematic offset of 0.07
dex in [Fe/H].

We assign a final systematic uncertainty of 0.13 dex to our [Fe/H]
distributions, adding the errors from continuum fitting and stellar
parameters in quadrature. This value is somewhat conservative, as much
of this uncertainty (particularly the continuum normalization) has
been calibrated out with the M13 data. We find, through simulations
similar to those previously described, that the observed [Fe/H]
scatter is well described by a combination of both Poisson noise and
statistical scatter of the underlying stellar parameters, again at the
level of $\pm100$K in $T_{\rm eff}$, $\pm0.3$ dex
in both $v_t$ and log$g$. Thus, summing these sources of scatter in
quadrature, we find for typical S/N and $\mbox{[Fe/H]} = -1$ a single-star
statistical error of 0.16 dex.

\subsection{Decomposing the [Fe/H] Distributions}\label{sec:decomp}

To characterize the Monoceros MDF, we must first decompose the
distribution of reliable [Fe/H] estimates in each field into Galactic
components (halo, disk) and Monoceros members. To estimate the
expected background in both halo and disk stars, we simulate CMDs for
each pointing with both the Besan\c{c}on model \citep{robin03} and
{\it galfast}, using the best fit Galactic parameters of
\cite{juric08}. All simulations included realistic photometry errors
and were extinction free, for comparison to observed SDSS CMDs
dereddened according to \cite{SFD}, henceforth SFD. We created a
matched filter in target density for the stars with reliable [Fe/H] in
each field, then integrated against the observed, Besan\c{c}on and
{\it galfast} star counts to determine the expected number of
contaminating stars from the disk and halo. For each field, overall
CMD normalization constants close to unity were applied by requiring
agreement between the total number of stars redder than $g-r=0.8$ and
falling within the targeted range of $r$ values. This analysis
suggests $\sim$65\% of Monoceros candidates with well-measured [Fe/H]
are true stream members. In no case do we expect any thin disk stars,
as the Monoceros targets have $|Z| \geq 2.5$ kpc.

\begin{figure*} 
 \begin{center} 
  \epsfig{file=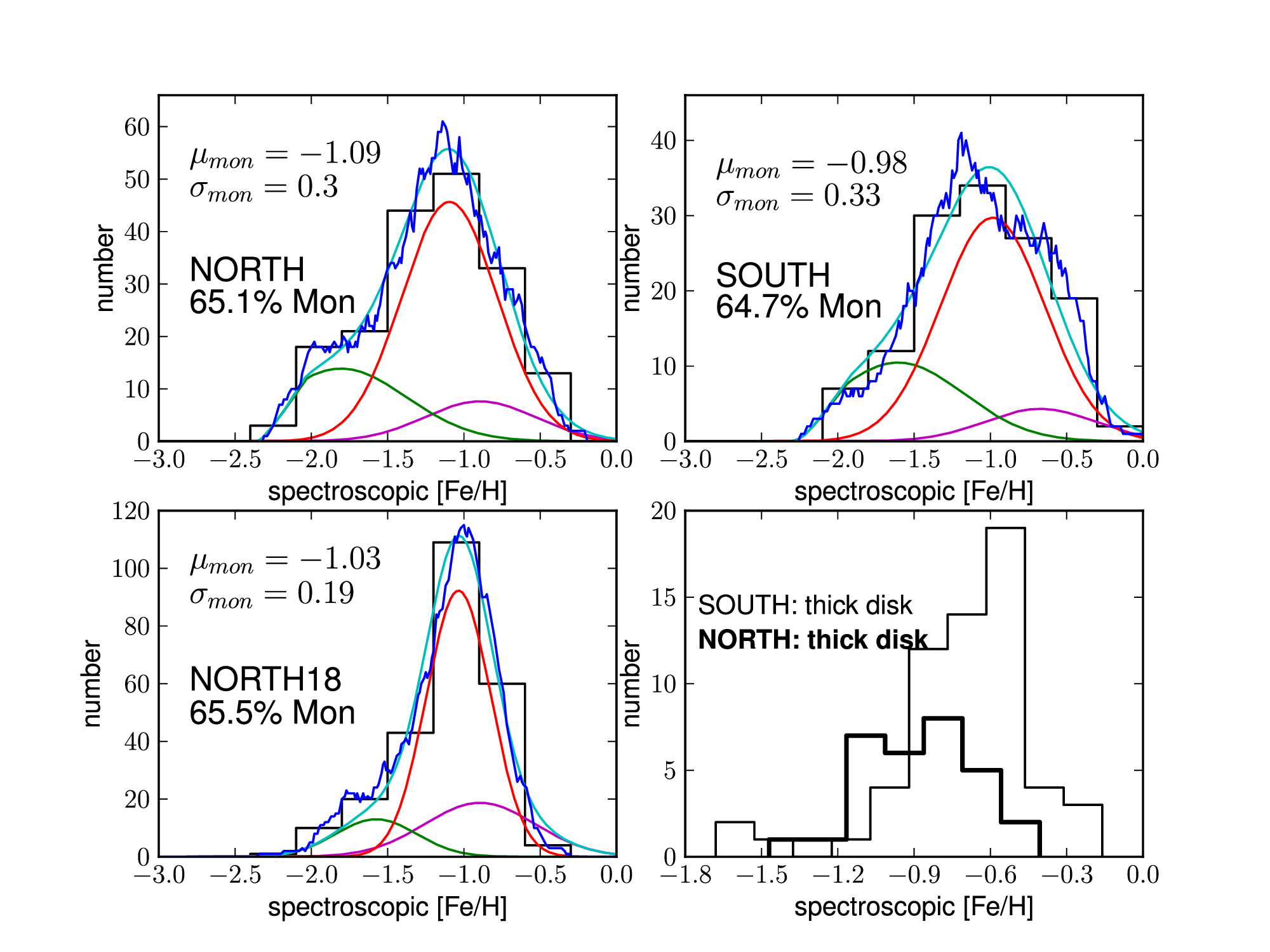, height=4.8in,width=6.0in} 
  \caption{\label{decomp} (\textit{top row, bottom left})
  Decompositions of the [Fe/H]$_{spec}$ distributions for each of our
  three Monoceros target fields. In each case, the green line
  represents the assumed halo contribution, the purple line represents
  the assumed thick disk contribution, the red line represents the
  best fit Monoceros contribution, and the cyan line gives the sum of
  these three components. The black histogram shows the observed
  [Fe/H]$_{spec}$ distribution in bins 0.3 dex wide, while the blue
  line is a curve which counts the number of objects within a 0.3 dex
  wide [Fe/H] bin, as the bin center is shifted incrementally by 0.01
  dex at a time. (\textit{bottom right}) [Fe/H]$_{spec}$ distributions
  of thick disk targets in Galactic north (thick black line) vs. south
  (thin black line) displaying the observed 0.22 dex metallicity
  asymmetry.}
 \end{center} 
\end{figure*}

Figure \ref{decomp} shows a decomposition of the observed
distributions of reliable [Fe/H] for each field. In principle, there
are nine free parameters: the normalization of Monoceros, thick disk,
and halo components, their median metallicities and their
dispersions. The following subsections describe our assumptions for
the thick disk and halo distributions, and procedure for fitting the
Monoceros component.

\subsubsection{Thick Disk Parameters}
In each field, the thick disk normalization was fixed by the CMD
analysis to be the average of the predictions from the {\it galfast}
and Besan\c{c}on simulations. In fields NORTH, SOUTH we fix the median
metallicity to that of the thick disk targets in these fields, plus a
small $-0.02$ dex correction from \cite{bond10} equation (A2) to
account for the vertical disk metallicity gradient (our thick disk
targets are at lower $|Z|$ than the more distant thick disk stars
which contaminate the Monoceros sample). For field NORTH18, in which
we did not target any thick disk stars, the median [Fe/H] used for
field NORTH was adopted. A $0.35$ dex intrinsic dispersion calculated
from the kinematically selected thick disk stars in \cite{bensby07}
was adopted for the thick disk in all fields.

\subsubsection{Halo Parameters}
In fields NORTH, NORTH18, the halo normalization was fixed to the
average of the {\it galfast} and Besan\c{c}on predictions. A cut on
$|v_{rad}|>150$ km~s$^{-1}$, yielding a subsample which should be
dominated by large velocity dispersion halo members, suggests that the
absolute number of halo stars in each field is roughly
constant. However, in field SOUTH, {\it galfast} gives an unreasonably
large halo contribution relative to this expectation, so in that case
we simply use the Besan\c{c}on value. For field NORTH18, the halo
median/dispersion were fit by examining the top quartile of stars in
S/N (62 stars, S/N$>96$), which displayed a relatively narrow peak at
$\mbox{[Fe/H]} = -1.57$. For field NORTH, we used SDSS proper motions from
\cite{munn} to isolate 7 halo stars with the cut $|v_{rad}|>100$
km~s$^{-1}$ and $\mu_{l}>220$ km~s$^{-1}$. This yielded a median
metallicity of $-$1.80. For field SOUTH, the kinematic data were of
lower quality, and we adopt $\mbox{[Fe/H]} = -1.6$ from \cite{carollo}. For
fields NORTH, SOUTH the dispersion was calculated by adding the errors
due to Poisson noise and stellar parameter scatter in quadrature with
a 0.3 dex intrinsic halo [Fe/H] dispersion (I08).

\subsubsection{Fitting Monoceros Parameters}
The normalization of the Monoceros distribution is fixed by requiring
that the components sum to the total number of targets. This leaves
two free parameters characterizing the Monoceros MDF: its median and
dispersion. All components were assumed to be intrinsically
gaussian. To account for detection inefficiency at very low [Fe/H], we
modulated the distributions by a linear ramp defined to be zero at
(and below) the minimum detected [Fe/H] and unity at (and above) the
maximum rejected [Fe/H]. For each field, we fit the Monoceros [Fe/H]
median $\mu_{mon}$ and dispersion $\sigma_{mon}$ by maximizing the
likelihood:

\begin{equation}
L=\frac{1}{n}\sum_{i=1}^{n} log_{10}\{P(\rm [Fe/H] \it _i;\mu_{mon},\sigma_{mon})\}
\end{equation}

\noindent
Where $n$ is the total number of reliable [Fe/H] measurements in each
field and $P(\rm [Fe/H])$ is the normalized sum of the halo, thick disk
and trial Monoceros modulated gaussian distributions.  Table
\ref{dist} lists the fitting assumptions and results, which agree with
expectations based on visual inspection.

\section{Calcium Abundance of the Monoceros Stream}\label{ab_ca} 

We have also measured [Ca/H] using the Ca\,II\,K line for program
stars sufficiently metal poor to avoid saturation effects
($\mbox{[Fe/H]} < -0.5$). Our [Ca/H] determination follows a procedure very
similar to that described for [Fe/H] in $\S$\ref{fe_fit}. For each
star, we adopt the same stellar parameters as in the [Fe/H] fits, but
now input the measured [Fe/H] as the Kurucz atmosphere metallicity. To
determine each star's [Ca/H] value, we minimize the same $\chi^2$
statistic, this time for $3922 \textrm{\AA} < \lambda < 3945
\textrm{\AA}$. The line list for Ca\,II\,K syntheses is provided in
Table \ref{linelist}.

\subsection{Continuum Normalization \& Calcium Calibration} \label{ca_cont}
For the Ca fitting region, we used a local continuum normalization
procedure, iteratively fitting a second order polynomial to the region
$3845 \textrm{\AA} < \lambda < 4045 \textrm{\AA}$. We found that this
continuum essentially matched the values/slopes of the continuum on
the two ``sideband'' regions blueward/redward of the Ca\,II~H/K
absorption. This fits the continuum shape well, but underestimates the
normalization by neglecting absorption which causes the continuum to
be rarely if ever realized. For this reason, these continuum fits
caused [Ca/H] to be systematically underestimated, with the effect
becoming more pronounced with increasing [Fe/H].

We recalibrate the continuum by requiring that our thick disk sample
exhibit a trend of [Ca/Fe] vs. [Fe/H] matching that from the
literature. The expected trend was computed using 150 Milky Way field
stars from \cite{gratton03}. We found good agreement in slope could be
achieved by mulitplying the initial continua by a factor of
($1+\delta$), with $\delta$ increasing linearly from zero at low
metallicity, $\mbox{[Fe/H]}_{\rm M13} = -1.55$, to 13\% at
$\mbox{[Fe/H]} = -0.50$. Agreement in offset was achieved with subsequent
addition of 0.03 dex to all [Ca/Fe]. Applying this procedure to our
189 M13 spectra yielded $\mbox{[Ca/Fe]} = 0.25$, in agreement with the estimate
of \cite{cohen}, who find $\mbox{[Ca/Fe]} = 0.19\pm0.07$ from high-resolution
spectroscopy of 25 cluster members.

\begin{figure*} 
 \begin{center} 
  \epsfig{file=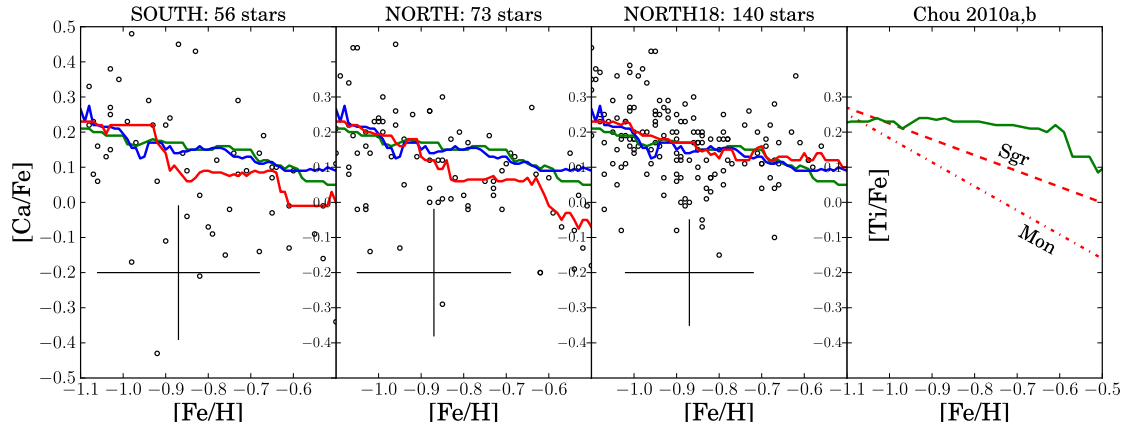, height=2.3in,width=6.9in} 
  \caption{\label{ca_comparison} (\textit{leftmost three panels})
  [Ca/Fe] as a function of [Fe/H] for Monoceros fields. The green line
  is the Galactic trend from \cite{gratton03}, the blue line
  represents our thick disk sample, and the red line represents
  Monoceros candidates. All lines are constructed by taking the median
  of individual measurements in bins 0.2 dex wide in [Fe/H] centered
  about the abscissa. Open circles are measurements for individual
  stars, and the crosses indicate typical single-star errorbars at the
  low metallicity end, $\mbox{[Fe/H]} = -1.1$. (\textit{rightmost panel}) The
  green line shows the [Ti/Fe] trend inferred from \cite{gratton03} in
  a fashion identical to that used to construct the green [Ca/Fe]
  lines at left.  The dot-dashed and dashed red lines give linear fits
  to [Ti/Fe] measurements of 21 Monoceros members and 70 Sagittarius
  stream members, respectively \citep{chou_sgr,chou_mon}. Two of three
  Monoceros fields (NORTH, SOUTH) show [Ca/Fe] underabundance trends
  relative to the Milky Way similar to those displayed by [Ti/Fe] in
  Sgr and Mon.}
 \end{center} 
\end{figure*}

We apply the local continuum fit and subsequent recalibration to
Monoceros candidates with $-1.1 < \mbox{[Fe/H]} < -0.5$, where we had
sufficient disk stars to verify the recalibration procedure. Figure
\ref{ca_comparison} overplots these Monoceros results with the
literature trend and recalibrated disk, where the lines are
constructed by taking the median of [Ca/Fe] in bins 0.2 dex wide in
[Fe/H]. Also plotted are linear fits to recent [Ti/Fe] abundances of
70 Sagittarius and 21 Monoceros stream members
\citep{chou_sgr,chou_mon}. The $\alpha$-enhancements [Ti/Fe] and
[Ca/Fe] are expected to display similar trends, as both reflect the
relative heavy element contributions of Type II
($\alpha$-elements$+$Fe) vs. Type Ia supernovae (Fe only), and hence
probe the star-formation history. dSph galaxies typically show
deficiencies in $\alpha$-elements relative to the Milky Way, owing to
a slower star formation rate in which fewer Type II explosions occur
before the onset of Type Ia explosions at $\ge$1 Gyr.

 \begin{deluxetable*}{cccccccccccc}
 \tabletypesize{\scriptsize}
 \tablecolumns{12} 
 \tablewidth{0pc} 
 \tablecaption{\label{dist} [Fe/H] Distribution Fitting Assumptions/Results} 
 \tablehead{
\multicolumn{5}{c}{} & \multicolumn{6}{c}{[Fe/H]} & \colhead{} \\ 
\cline{6-11} 
 \colhead{Field} & 
 \colhead{N$_{stars}$} &
 \colhead{N$_{halo}$} & 
 \colhead{N$_{TD}$} & 
 \colhead{N$_{mon}$} & 
 \colhead{$\mu_{TD}$} & 
 \colhead{$\sigma_{TD}$} (dex) & 
 \colhead{$\mu_{halo}$}       & 
 \colhead{$\sigma_{halo}$} (dex) & 
 \colhead{$\mu_{mon}$} & 
 \colhead{$\sigma_{mon}$ (dex)} & 
 \colhead{$L_{max}$}
 }
 \startdata 
  NORTH   & 183 & 40.8 & 23.2 & 119.2 & $-0.89$ & 0.35 & $-$1.80 & 0.42 & $-1.09$ & $0.30$ & $-0.21$ \\
  SOUTH   & 131 & 33.3 & 13.0 & 84.7  & $-0.67$ & 0.35 & $-$1.60 & 0.42 & $-0.98$ & $0.33$ & $-0.23$ \\
  NORTH18 & 247 & 28.7 & 56.4 & 161.8 & $-0.89$ & 0.35 & $-$1.57 & 0.25 & $-1.03$ & $0.19$ & $-0.10$ 
\enddata
 \end{deluxetable*}

\vspace{0.2in}

\section{Discussion}\label{disc}
\subsection{Monoceros Abundances}
As the Monoceros members comprise $\sim$2/3 of targets in each field,
it can be seen by visual inspection that Monoceros has a median
$\mbox{[Fe/H]} \sim -1$, distinct from both the thick disk and halo. Indeed,
our maximum likelihood fits yield $\mbox{[Fe/H]} \sim -1.0$ for each
subsample, showing no trend with $l$ (see Table \ref{dist}). I08
analyzed $\sim$$11,000$ MSTO stars in a similar spatial region to that
of our targets ($13<R_{\rm GC}$/kpc $<16$,
$3<Z$/kpc $<4$, $170^{\circ} < l < 190^{\circ}$, see their Figure 18);
our inferred Monoceros medians $-1.09 < \mbox{[Fe/H]} < -0.98$ are consistent
with the I08 photometric estimate of $\mbox{[Fe/H]} = -0.95$ to within the
systematic uncertainty of $\sim$0.13 dex on our spectroscopic
abundances.  It should be noted that we have used the same $g-r$
temperature calibration as I08. A more stringent test of SDSS
photometric metallicities would be provided by a study deriving
$T_{\rm eff}$ independently. See
$\S$\ref{sec:phot} for a detailed comparison of photometric
vs. spectroscopic metallicities.

I08 also inferred a small intrinsic Monoceros [Fe/H] dispersion of
only $0.15$ dex. For each field, we simulated the distributions of
[Fe/H] arising from a monometallic population at $\mbox{[Fe/H]} = \mu_{mon}$,
given our S/N distribution, $T_{\rm eff}$
distribution, continuum placement errors and expected stellar
parameter scatter. We found RMS measurement-induced scatter of 0.21
dex, 0.23 dex, and 0.16 dex for fields NORTH, SOUTH, NORTH18, implying
intrinsic scatters of 0.21 dex, 0.22 dex, 0.10 dex
respectively. Applying the same methodology to M13 stars of comparable
S/N ($>$29), we find a measurement error of 0.36 dex, and observed
scatter of 0.42 dex, yielding an intrinsic M13 scatter of 0.20
dex. Globular clusters generally have RMS iron abundance dispersion
$\le 0.05$ dex \citep{carretta}. However, with only a single exposure,
the M13 sample is subject to additional sources of error (e.g., cosmic
rays) which we have not modeled, perhaps reconciling this
difference. If we are indeed slightly underestimating the
observation-induced scatter, a very conservative interpretation would
regard the intrinsic Monoceros dispersions we obtained as upper
limits.

For fields NORTH, NORTH18, a single gaussian component is sufficient
to describe the Monoceros MDF. For field SOUTH, there appears to be an
excess at $\mbox{[Fe/H]} \sim -1.2$. This could arise from a multimodality in
the Monoceros MDF or, more likely, indicates contamination by the
metal weak thick disk (MWTD, see $\S$\ref{MWTD}). Inspection of the
CMD in field SOUTH also suggests multiple main sequence turnoffs aside
from that of the disk. Given these uncertainties in population
modeling, the inferred gradient in Monoceros median metallicity with
$b$ ($+0.11$ dex from $b=+25^{\circ}$ to $b=-25^{\circ}$) appears not
to be significant.

In two of three fields, the Monoceros [Ca/Fe] dips below the Galactic
trend as [Fe/H] increases from $-1.1$ to $-0.5$ (see Figure 3). In
fields NORTH and SOUTH, this [Ca/Fe] trend is qualitatively similar to
the recently measured [Ti/Fe] trends of Monoceros and Sagittarius
stream members \citep{chou_mon,chou_sgr}. In field NORTH18, the
observed Monoceros trend appears to exactly match that of the Milky
Way. However, the field NORTH18 decomposition in Figure \ref{decomp}
suggests the measured [Fe/H] distribution cuts off more sharply than
expected at high metallicity. If indeed [Fe/H] has been systematically
underestimated for the highest metallicity stars in this field,
[Ca/Fe] would be artificially inflated. It should be noted that our
Monoceros sample plotted is a mixture of Monoceros and disk stars,
driving down the apparent contrast between the Monoceros and Milky Way
trends.

\subsection{Thick Disk Abundances} \label{thickdisk}

In addition to the Monoceros candidates, in fields NORTH and SOUTH we
measured [Fe/H] for 30 and 60 thick disk stars respectively. These
fields exhibit a strong stellar number density asymmetry, with the
southern field showing an excess of more than 50\% compared to its
northern counterpart. The asymmetry is easily detected in the SDSS
data, is also detectable in Pan-STARRS1 $3\pi$ survey data, and cannot
be accounted for by photometric errors or reddening
uncertainties. Juri{\'c} et al. (in preparation) speculate it may be a
signature of yet another, more nearby, stream, or otherwise a global
disturbance (a ``warp'' or a ``bend'') in the thick disk, of uncertain
origin.

Figure \ref{decomp} shows the measured distributions; the median and
dispersion of [Fe/H] in field SOUTH are $-$0.65, 0.29 dex
respectively, while for field NORTH, the corresponding values are
$-$0.87, 0.22 dex. The statistical scatter in our [Fe/H] measurments $\sim$0.05 dex per star has negligible effect on these values. The median $|Z|$ for the NORTH, SOUTH thick disk
targets is respectively 1.60 kpc and 1.79 kpc, using the photometric
parallax relation of I08 equations (A6) and (A7). According to the
gradient in disk metallicity from \cite{bond10} this corresponds to an
expectation, given perfect disk symmetry, that SOUTH would have [Fe/H]
0.01 dex \textit{lower} than NORTH. The asymmetry is not simply
reflecting different scale heights. Although the adopted systematic
error on our metallicities is nominally 0.13 dex, this represents an
offset with respect to the true abundances; our internal consistency
should be much better, especially for these two thick disk samples
which have very similar colors and S/N, as well as relatively strong
Fe features. Reconciling the median metallicities of NORTH/SOUTH thick
disk sample would require that stellar parameters such as log$g$ and
$v_t$ differ greatly between the two populations, which itself would
require invoking disk asymmetry about the midplane. A
Kolmogorov-Smirnov test between the two measured [Fe/H] distributions
yields a consistency probablility of $<$$2.5\times10^{-3}$, even
excluding the added 0.01 dex contrast from the $|Z|$ gradient in
[Fe/H].

The dispersion in [Fe/H] for field SOUTH is 0.22 dex, versus 0.29 dex for field NORTH and 0.35 dex for the thick disk (averaged over many directions)
according to \cite{bensby07}. A disrupted dSph origin for the overdensity would naturally explain the lower [Fe/H] dispersion in field SOUTH, but also incorrectly predict an [Fe/H] lower than that of the northern disk for the southern stellar overdensity. 

Thus, we have spectroscopically identified an asymmetry in the thick
disk towards the Galactic anticenter, at
$R_{\rm GC} \approx 12$ kpc and $|Z| \approx 1.7$
kpc, which we expect to be confirmed and characterized further via
stellar counts analysis.

\subsection{Comparison with Photometric Metallicities}\label{sec:phot}
 
With $ugriz$ photometry available, we can compute [Fe/H]$_{phot}$ from
equation (A1) of \cite{bond10} for the entire sample. [Fe/H]$_{phot}$
is reddening-sensitive. \cite{schlafly10} have recalibrated SFD,
finding the need for a substantial, $\sim$45\%, global change in
$E(u-g)$, but only a $2$\% change in $E(g-r)$. Thus, our
$T_{\rm eff}$ are negligibly affected at a
$\lesssim$10K level. But for fixed $g-r$, and near $\mbox{[Fe/H]}_{phot} = -1$,
[Fe/H]$_{phot}$ has a strong $u-g$ gradient, $>$0.04 dex per 0.01 mag
$u-g$. The \cite{schlafly10} recalibration implies $\Delta(u-g)>$0.06
in portions of our sample, a major effect. However, the
[Fe/H]$_{phot}$ estimator of \cite{bond10} is tied to SFD reddenings,
and we therefore calculate [Fe/H]$_{phot}$ using $(u-g)_{SFD}$; it may
be necessary to revise the SDSS [Fe/H]$_{phot}$ relation using updated
$E(u-g)$ values if these photometric metallicities are to be
independent of dust column.

Furthermore, single-pointing samples of [Fe/H]$_{phot}$ and
[Fe/H]$_{spec}$ are sensitive to local reddening in differing ways
([Fe/H]$_{spec}$ is a function of $g-r$ reddening through
$T_{\rm eff}$). Near [Fe/H] = $-1$, using
conventional reddening laws, overestimating $A_r$ leads to an
underestimate of [Fe/H]$_{phot}$, but to an overestimate of
[Fe/H]$_{spec}$ for fixed absorption EW. Any local (degree scale)
error in $A_r$ drives our photometric and spectroscopic metallicities
apart.

Nevertheless, we find a highly significant correlation between
[Fe/H]$_{phot}$ and [Fe/H]$_{spec}$. The correlation is strongest for
the highest S/N spectra, corresponding to the brightest objects with
least noisy $ugr$ photometry and best measured [Fe/H]$_{spec}$. M13,
closely approximating a monometallic population, shows only a marginal
correlation, in accordance with the expectation that the observed
spread in [Fe/H] values arises predominantly from random scatter in
the photometric/spectroscopic measurements.

\begin{figure*} 
 \begin{center} 
  \epsfig{file=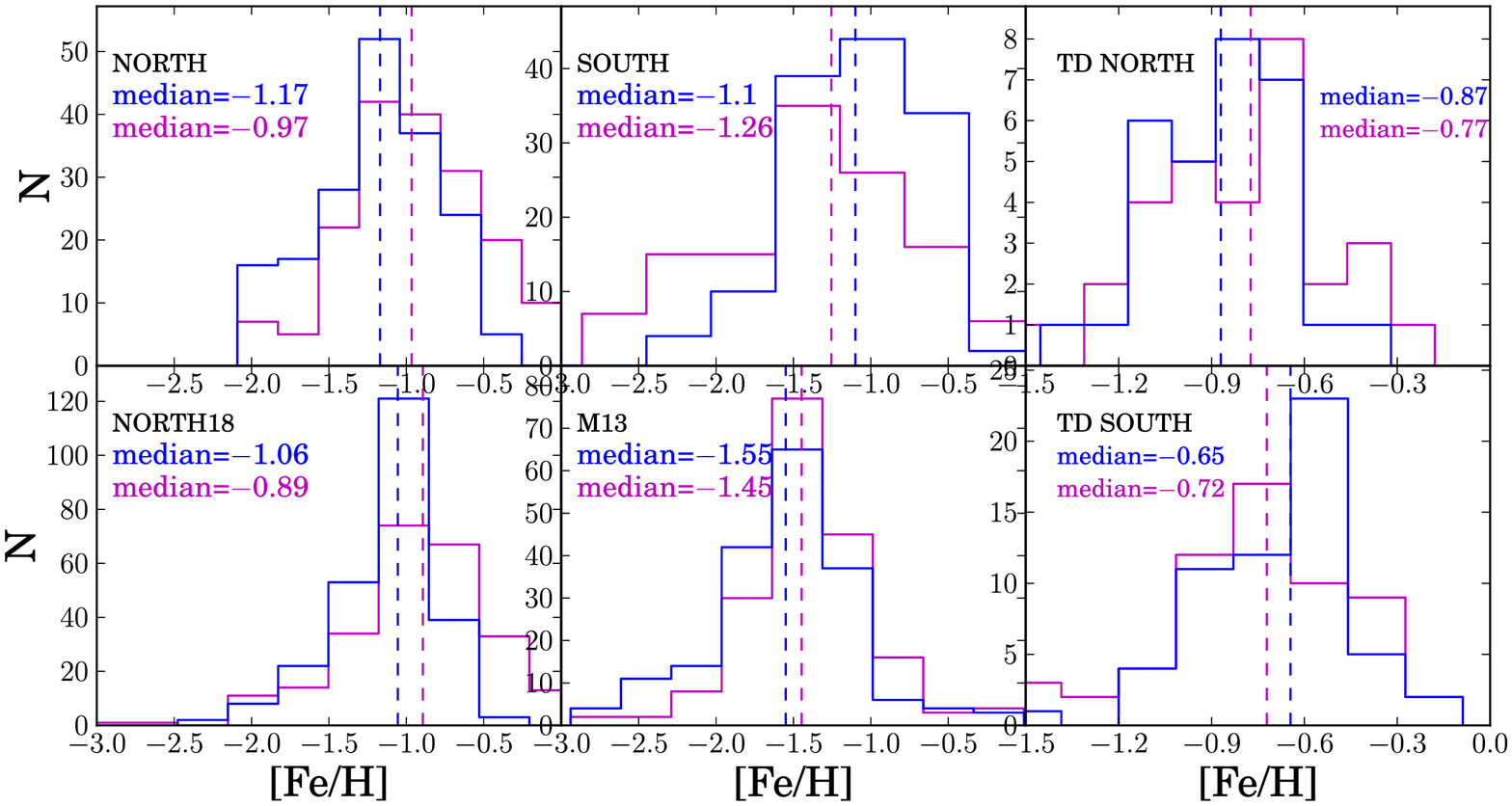,height=3.4in,width=6.8in} 
  \caption{\label{comparison} Photometric (magenta) vs. spectroscopic
  (blue) [Fe/H] distributions. The distributions generally agree at
  the level of $\sim$$0.15$ dex, which can be accounted for by the
  estimated $0.13$ dex systematic error on our spectroscopic
  metallicities and the 0.1 dex systematic error on
  [Fe/H]$_{phot}$. In general, the photometric metallicity
  distributions are slightly broadened relative to the spectroscopic
  distributions.}
 \end{center} 
\end{figure*}

Figure \ref{comparison} overplots the distributions of
photometric/spectroscopic metallicities for each of our subsamples. In
general, the median photometric and spectroscopic metallicities agree
at the level of $\sim$$0.15$ dex, which can be reconciled by the
estimated systematic uncertainties of 0.13 dex on [Fe/H]$_{spec}$ and
0.1 dex on [Fe/H]$_{phot}$ (I08). Except for the M13 sample, the
[Fe/H]$_{spec}$ distributions are narrower than their [Fe/H]$_{phot}$
counterparts, by a median of $0.07$ dex in dispersion. Of course, much
of the total dispersion is intrinsic. The superiority of M13
[Fe/H]$_{phot}$ owes to the targets being relatively bright
($r\sim18$), yielding good photometry, yet receiving only a single,
short Hectospec exposure.

\subsection{Radial Velocities}
As a byproduct of our abundance extraction procedure, we have heliocentric 
radial velocities in hand for our entire sample. Typical statistical errors are 
10-20 km~s$^{-1}$, depending on S/N. We gauged our systematic error using M13 as 
a calibration sample, since this globular cluster has a known $v_{rad}$ = $-$244.2 
km~s$^{-1}$ \citep{harris96}. We find for our 189 M13 spectra a median of $v_{rad}$ = $-$241.1 
km~s$^{-1}$, and therefore have applied a correction of $-3.1$ km~s$^{-1}$ to all radial 
velocity measurements.

\subsubsection{Monoceros Candidates}
\cite{yanny03} found that Monoceros is kinematically cold, consistent with the expectation for a disrupted dSph galaxy, measuring radial velocity dispersions 
$\sigma_r$ = 22-30 km~s$^{-1}$. For stars with $\mu_{mon}-\sigma_{mon} \le $ [Fe/H] $ \le \mu_{mon}+\sigma_{mon}$ in fields NORTH, SOUTH, NORTH18, we find 
$\sigma_r = $35.7, 31.3, 26.3  km~s$^{-1}$ after correcting for median measurement uncertainties of 17, 24, 11 km~s$^{-1}$ respectively. For comparison, 
the \textit{galfast} (Besan\c{c}on) predictions for $\sigma_r$ of thick disk stars in these fields is  57.7 (38.4), 56.6 (36.3), 54.3 (35.8) km~s$^{-1}$. All
dispersions were calculated as half of the difference between the 84th and 16th percentile $v_{rad} $ values. In all cases, the metallicity-selected Monoceros 
sample is kinematically cold relative to the thick disk model, and appreciably so if we to are adopt the \textit{galfast} prediction. Our range of $\sigma_r$ is 
slightly higher than that found by \cite{yanny03}, but this may be due to contamination of the sample by disk and halo stars, both of which would tend to
increase $\sigma_r$.

\begin{figure*} 
 \begin{center} 
  \epsfig{file=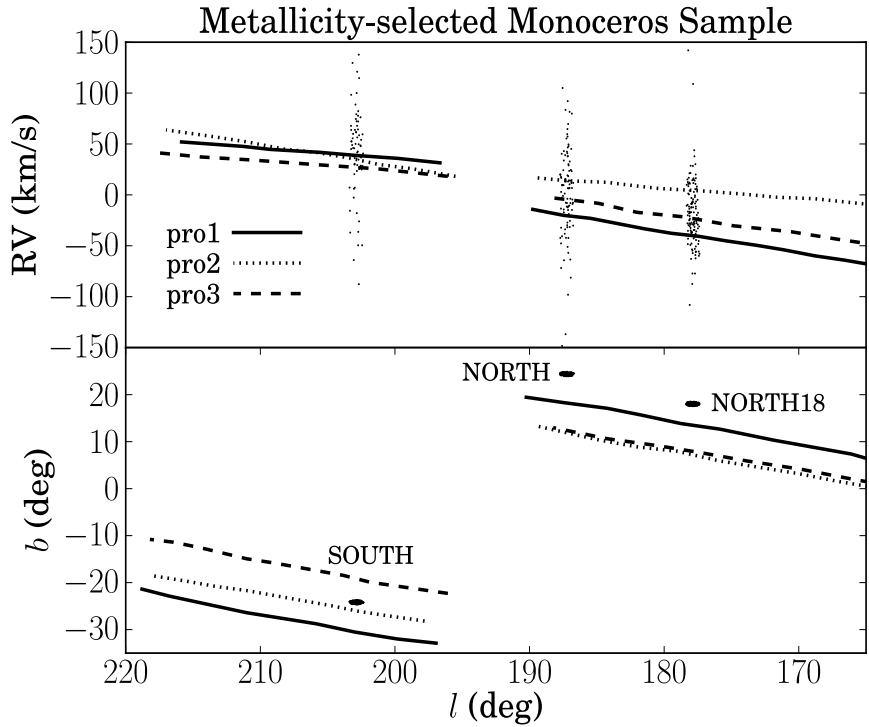,height=4in,width=5in} 
  \caption{\label{rv_longitude} Adaptation of \cite{penarrubia} Figure 7. (\textit{bottom}) Prograde orbit models (pro1, pro2, pro3)
 of Monoceros debris position in the anticenter 
direction. The gap at $l \sim 195^{\circ}$ signifies that field SOUTH corresponds to a different orbital ``wrap'' than NORTH, NORTH18. (\textit{top}) 
Model-predicted radial velocity as a function of galactic longitude, with each point representing a star in our metallicity-selected Monoceros sample.}
 \end{center} 
\end{figure*}

\begin{figure} 
% \begin{center} 
  \epsfig{file=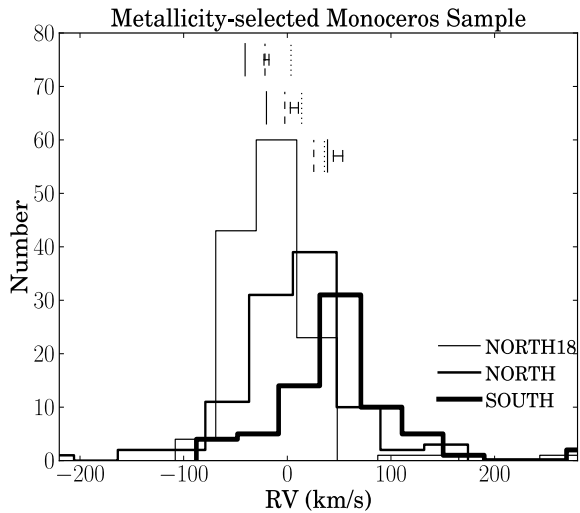,height=2.8in,width=3.5in} 
  \caption{\label{rv_hist} Histograms of metallicity-selected Monoceros candidate radial velocities by field. Capped, solid vertical lines represent measured
$v_{rad}$ medians and their standard errors. Corresponding (uncapped) vertical lines represent the model predictions at each field location, with the same linetype
legend as in Figure \ref{rv_longitude}.}
% \end{center} 
\end{figure}

We can also compare the median $v_{rad}$ for the metallicity-selected Monoceros samples to the predictions of \cite{penarrubia}. Figure \ref{rv_longitude} 
overplots measured radial velocities with the three prograde orbit models of \cite{penarrubia}. Figure \ref{rv_longitude} also plots $l$ vs. $b$ for our Monoceros 
targets and the \cite{penarrubia} orbits, showing that field SOUTH is on a different ``wrap'' of
the stream than are NORTH, NORTH18.There is indeed qualitative agreement between all of the models 
and our measured median $v_{rad}$ for NORTH, SOUTH, NORTH18 of 7 $\pm$ 2.9, 49 $\pm$ 4.5, $-$20 $\pm$ 2.4 km~s$^{-1}$, trending upwards with increasing $l$.
 Quantitative agreement is only achieved at the $\sim$10 km~s$^{-1}$ level, with no single model particularly favored, 
as shown by the histograms of the metallicity-selected Monoceros $v_{rad}$ measurements in each field (Figure \ref{rv_hist}). The median Monoceros $v_{rad}$ 
values generally differ substantially from the \textit{galfast} (Besan\c{c}on) thick disk $v_{rad}$ predictions of 30 (22), 54 (59), 5 (6) km~s$^{-1}$
for fields NORTH, SOUTH, NORTH18 respectively. In summary, the measured radial velocities of our metallicity-selected Monoceros subsamples are kinematically cold
relative to the thick disk, have different median values than expected for thick disk stars, and generally agree with the model predictions of \cite{penarrubia}.

%compare median values to TD values

%add comparison of v_rad to model predictions for HALO
%could add sentence stating we can't get clean thick disk sample to empirically get thick disk dispersion

\subsubsection{Thick Disk Targets}

Since, as discussed in $\S$\ref{thickdisk}, we find that the thick disk SOUTH sample is chemically distinct from its northern counterpart, it is 
of interest to compare the measured thick disk radial velocities with \textit{galfast}/Besan\c{c}on predictions. For both the NORTH and SOUTH thick disk samples, 
we find rather poor agreement in terms of median $v_{rad}$, with measured values of $-$8, 31 km~s$^{-1}$ versus 
\textit{galfast} (Besan\c{c}on) predicted values of 21 (18), 52 (49) km~s$^{-1}$. The measured dispersions $\sigma_r$ for NORTH, SOUTH are 42, 25 km~s$^{-1}$, versus \textit{galfast} (Besan\c{c}on) predicted values of 
46 (47), 46 (49) km~s$^{-1}$. The measured velocity dispersion in field NORTH agrees with the thick disk predictions, whereas the SOUTH targets are much more 
kinematically cold than expected. Thus, the thick disk kinematics also display a north/south asymmetry, with a low velocity dispersion for the southern overdensity 
suggestive of a coherent stream.

\section{Conclusions}\label{conclusions}  

\subsection{What is the Monoceros Iron Abundance?}

We have presented the first ever spectroscopic Monoceros MDF based
directly on Fe absorption line measurements, finding
$\mbox{[Fe/H]} = -1.0\pm0.1$. Our results confirm the photometric metallicity
analyses of I08 ($\mbox{[Fe/H]} = -0.95$) and \cite{sesar11}, who also used the
\cite{bond10} calibration to derive results consistent with
$\mbox{[Fe/H]} \sim -1.0$ towards
($l$,$b$)=(232$^{\circ}$,26$^{\circ}$). However, these earlier results
are vulnerable to systematic problems (e.g., $u$ band extinction
uncertainties) which are largely circumvented by our spectroscopic
analysis.

While C03 and Y03 inferred [Fe/H] for Monoceros spectroscopically,
neither measured Fe absorption directly. Instead, both relied upon
indices calibrated to Mg and Ca features. Our [Ca/Fe] results suggest
that such calibrations of [Fe/H] to standard $\alpha$-element trends
may not be justified. Further, in the case of Y03, whose technique
measured Ca\,II\,K EW, our analysis shows that Ca and Fe absorption in
this region are highly degenerate, implying large uncertainties for
the Y03 procedure. There does not appear to be a simple means of
bringing about agreement between our result $\mbox{[Fe/H]} = -1$ and that of
Y03, given the similar location of their targets at
($l$,$b$)=($198^{\circ}$,$-27^{\circ}$), $d_{\odot} = 13$ kpc and our
field SOUTH at ($l$,$b$)=($203^{\circ}$,$-24^{\circ}$), $d_{\odot}=12$
kpc. To whatever extent the present results disagree with those of C03
and Y03, our results should take precedence, as we have measured
[Fe/H] directly, analyzing spectra with depth-of-exposure
$\sim$7$\times$ that of Y03 and S/N comparable to that of C03. Our
sample size also eclipses that of C03 by a factor $>$10$\times$ and
that of Y03 by a factor $\sim$1.5$\times$.

Assuming the Monoceros progenitor contained stellar populations of
varying age, we can still reconcile our result with that of the C03
M-giant sample, $\mbox{[Fe/H]} = -0.4\pm0.3$. M-giants preferentially trace
younger, higher metallicity populations than MSTO
stars. \cite{carlin11} have noted that this bias reconciles their MSTO
Sgr stream $\mbox{[Fe/H]} = -1.15$ with literature M-giant values of
$\mbox{[Fe/H]} \sim -0.6$. The Monoceros case appears directly analogous
given our conclusion that Monoceros has MSTO $\mbox{[Fe/H]} = -1.0$. This
reasoning can also bring about agreement between our $\mbox{[Fe/H]} = -1$ and
the median $\mbox{[Fe/H]} = -0.71$ of the 21 kinematically-selected M-giants
studied by \cite{chou_mon}.

Since we have measured [Fe/H] directly from the Fe absorption lines of
MSTO stars (a population with relatively little metallicity bias) and
with S/N and sample size comparable to or better than previous
spectroscopic efforts, we recommend adoption of our value
$\mbox{[Fe/H]} = -1.0\pm0.1$ as the metallicity of the Monoceros stream.

\subsection{The Metal Weak Thick Disk} \label{MWTD}

\cite{carollo} have speculated that the MWTD and Monoceros may share a
common origin. This notion is based in part on the coincidence between
the preliminary \cite{wilhelm05} estimate of Monoceros' metallicity,
$\mbox{[Fe/H]} = -1.37$, and their inferred
$\langle\mbox{[Fe/H]}\rangle_{MWTD} = -1.3$. Our measurements argue against
this MWTD-Monoceros connection, as we find a $\sim$0.3 dex offset in
[Fe/H] between these populations. All of our systematics are
characteristically $\sim$0.05-0.10 dex, and crudely speaking we rule
out a median $\mbox{[Fe/H]} = -1.3$ at the 2-3$\sigma_{syst}$ level. Decreasing
our median metallicity to $\mbox{[Fe/H]}_{spec} = -1.3$ would require, for
example, an unreasonably large temperature scale miscalibration of
$\sim$330K.

\begin{figure*} 
 \begin{center} 
  \epsfig{file=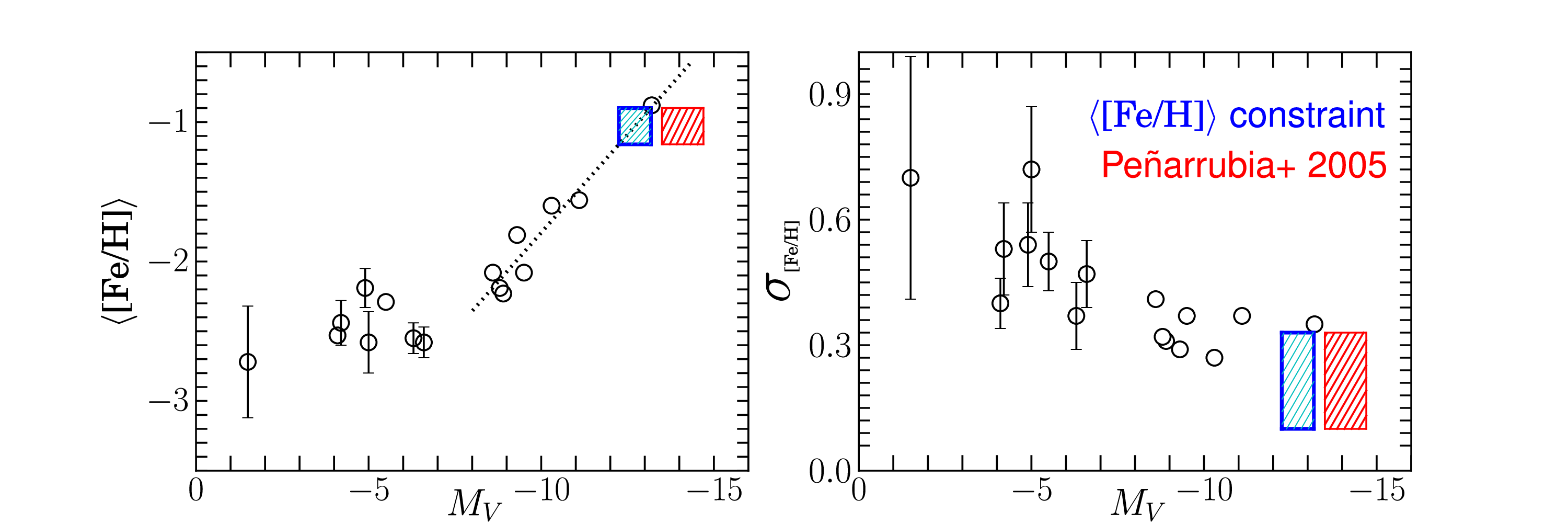,height=2.0in,width=6.0in} 
  \caption{\label{dwarf} (\textit{left}) $\langle$[Fe/H]$\rangle$
  vs. $M_V$, where each circle represents a single dwarf galaxy, with
  properties quoted from \cite{norris10} and referefences
  therein. Errorbars are only shown when the uncertainty on
  $\langle$[Fe/H]$\rangle$ exceeds 0.1 dex. The dashed line shows our
  linear fit to the eight plotted galaxies brighter than
  $M_V$=$-8$. The blue rectangle represents the allowed region of
  parameter space, combining this computed trend with our measured
  MDF, and constrains the Monoceros progenitor to have $-13.2 \leq M_V
  \leq -12.3$. Alternatively, we can use the Monoceros progenitor mass
  estimated by \cite{penarrubia} and assume the Sgr mass-to-light
  ratio to plot our Monoceros MDF (red hatched rectangle). Clearly,
  Monoceros follows the general dwarf galaxy trend of increasing
  $\langle$[Fe/H]$\rangle$ with increasing
  luminosity. (\textit{right}) $\sigma_{\rm [Fe/H]}$
  vs. $M_V$. Circles are individual dwarf galaxies with properties
  drawn from \cite{norris10} and references therein. Errorbars are
  only shown when the uncertainty on $\sigma_{\rm [Fe/H]}$
  exceeds 0.05 dex. Again, we have plotted Monoceros as the red
  hatched rectangle by assuming the \cite{penarrubia} mass and Sgr
  mass-to-light ratio, and as a blue hatched rectangle using the
  $\langle$[Fe/H]$\rangle$ constraint on $M_V$. Monoceros also follows
  the dwarf galaxy trend of decreasing
  $\sigma_{\rm [Fe/H]}$ with increasing luminosity.}
 \end{center} 
\end{figure*}

\subsection{The Monoceros Progenitor}

Much controversy and speculation has surrounded the origin of the
Monoceros structure since its discovery. To characterize the stream's
nature, Y03 used star counts to estimate the total stellar mass of
Monoceros at $\sim$$2\times10^{7} M_{\odot}$-$5\times10^{8}
M_{\odot}$, a range consistent with the content of a relatively
luminous dwarf galaxy. \cite{penarrubia} specifically simulated the
case of a dSph progenitor, fitting observations well with an initital
progenitor of total mass 3-9$\times10^8 M_{\odot}$. Still, detractors
\citep[e.g.,][]{momany} argue that there is no need for an
extragalactic origin of Monoceros. How can our observations further
constrain the nature of the Monoceros progenitor?

The trends of $\langle$[Fe/H]$\rangle$ and its dispersion,
$\sigma_{\rm [Fe/H]}$, with dwarf galaxy
luminosity have been studied by \cite{norris10}, and are reproduced in
Figure \ref{dwarf}. Dwarf galaxy $\langle$[Fe/H]$\rangle$ increases
monotonically with luminosity; by fitting a linear trend in
$\langle$[Fe/H]$\rangle$ vs. $M_V$ for
$-2.1 < \langle\mbox{[Fe/H]}\rangle < -0.9$, we can translate our MDF
measurement $\langle\mbox{[Fe/H]}\rangle \approx -1.0 \pm 0.1$ into a
constraint on the Monoceros progenitor,
$-13.2$$\leq$$M_V$$\leq$$-12.3$, assuming the progenitor was a dwarf
galaxy (see the dashed line and blue rectangle in Figure
\ref{dwarf}). Using a typical stellar mass-to-light ratio for dwarf
galaxies $(M/L_{V})_{*}$$\approx$3 \citep{chilin},
$\langle\mbox{[Fe/H]}\rangle=-1$ implies $3.3 \times 10^{7}$$M_{\odot}$
of stellar mass in the Monoceros progenitor. Provided the fraction of
the Monoceros progenitor which has been stripped is similar to that of
Sgr \citep[$>$2/3,][]{lm10}, we conclude that our estimate of the
stellar mass in the Monoceros stream based on chemistry and the
assumption of a dwarf galaxy origin is consistent with the independent
Y03 estimate from observed star counts. Having constrained the
progenitor luminosity, we can also place Monoceros on the Figure
\ref{dwarf} plot of $M_V$
vs. $\sigma_{\rm [Fe/H]}$, finding very good
qualititave agreement between Monoceros (blue rectangle) and other
bright, low metallicity dispersion dwarf galaxies.

Instead of deriving $M_V$ from $\langle$[Fe/H]$\rangle$, we can check
whether a dSph progenitor with best-fitting mass determined by
\cite{penarrubia} conforms the \cite{norris10} MDF trends. Taking the
Monoceros stream to have the same total mass-to-light ratio as the Sgr
core \citep[$M_V$=$-13.3$,
$M_{tot}$=$2.5\times10^8$$M_{\odot}$,][]{majewski03,lm10}, the
\cite{penarrubia} mass range translates to
$-14.7$$<$$M_V$$<$$-13.5$. We have combined this constraint with our
measured MDF to plot the \cite{penarrubia} Monoceros progenitor (red
rectangles) in Figure \ref{dwarf}. In both cases, $M_V$
vs. $\langle$[Fe/H]$\rangle$ and $M_V$
vs. $\sigma_{\rm [Fe/H]}$, Monoceros shows
qualitative agreement with the dwarf galaxy trend.

Thus, we conclude that our abundance study supports a dwarf galaxy
origin of Monoceros, in that: (1) The MDF median and dispersion both
conform to the expectations for an appropriately luminous dwarf galaxy
progenitor, and (2) We have detected [Ca/Fe] deficiencies typical of
dwarf galaxy stellar populations in two of three fields observed.

\subsection{Future Extensions}

In the future, similar observations/analysis could be applied to
Monoceros fields that more optimally search for $l$ and $b$ abundance
gradients in the stream. For example, fixing $l$ and observing two
$|Z|$ values symmetrically above/below the Galactic midplane would
better isolate Monoceros abundance trends with $b$. Detecting or
ruling out a gradient toward the midplane in this way would
significantly constrain models which claim the stream to be a disk
``warp" or ``flare" \citep[e.g.,][]{momany}. Observing an extended
range of $l$ values at fixed $b$ would better constrain a metallicity
gradient along the stream. For example, Sgr exhibits such a gradient,
but only at the level $\sim$2.4$\times 10^{-3}$ dex per degree along
the debris trail \citep{keller}. To probe a gradient at this level
with spectroscopy comparable to that reported here, observations would
need to span $\sim$100$^{\circ}$ in $l$, a factor of four increase
relative to our current analysis, $178^{\circ}\leq l \leq
203^{\circ}$.

\acknowledgments We warmly thank Nelson Caldwell for his advice on MMT
 observations, Hectospec details and queue scheduling, as well as Evan
 Kirby and John Norris for fruitful discussions. We furthermore thank
 the queue observers and the SAO Telescope Data Center for reducing
 the data.  This research made extensive use of the Vienna Atomic Line
 Database (VALD). A.~M.~M. is supported by a National Defense Science
 \& Engineering Graduate fellowship. A.~F. is supported by a Clay
 Fellowship administered by the Smithsonian Astrophysical
 Observatory. M.~J. acknowledges support by NASA through Hubble
 Fellowship grant \#HF-51255.01-A awarded by the Space Telescope
 Science Institute, which is operated by the Association of
 Universities for Research in Astronomy, Inc., for NASA, under
 contract NAS 5-26555.

{\it Facilities:} \facility{MMT(Hectospec)}

 \begin{deluxetable}{cccccccccccc}
 \tabletypesize{\scriptsize}
 \tablecolumns{12} 
 \tablewidth{0pc} 
 \tablecaption{\label{linelist} Fe $\lambda$5269, Ca\,II\,K Line Lists} 
 \tablehead{
 \multicolumn{4}{c}{Fe $\lambda$5269} & \multicolumn{4}{c}{Ca\,II\,K} & \multicolumn{4}{c}{Ca\,II\,K cont'd} \\ 
 \cline{2-3}
 \cline{6-7}
 \cline{10-11}
 \colhead{$\lambda \ (\textrm{\AA})$} & \colhead{Species} & \colhead{E.P. (eV)} & \colhead{log$gf$} & \colhead{$\lambda \ (\textrm{\AA})$} & \colhead{Species} & \colhead{E.P. (eV)} & \colhead{log$gf$} & \colhead{$\lambda$ \ (\textrm{\AA})} & \colhead{Species} & \colhead{E.P. (eV)} & \colhead{log$gf$}
 }
 \startdata
   5268.6 & Ti\,II & 2.59 & $-1.96$ &  3922.1 & Fe\,I & 3.29 & $-2.44$ &  3933.3 & Sc\,I & 2.3 & $-1.29$ \\
  5268.9 & Ti\,I & 3.32 & $-1.74$ &  3922.7 & Fe\,I & 2.99 & $-2.0$ &  3933.4 & Sc\,I & 0.02 & $-0.65$ \\
  5269.5 & Fe\,I & 0.86 & $-1.32$ &  3922.9 & Fe\,I & 0.05 & $-1.65$ &  3933.6 & Fe\,I & 3.07 & $-1.16$ \\
  5269.9 & Ti\,I & 1.87 & $-1.74$ &  3923.0 & Sc\,I & 1.99 & $-2.69$ &  3933.6 & Fe\,I & 3.27 & $-2.05$ \\
  5270.3 & Ca\,I & 2.52 & $0.16$ &  3923.0 & Fe\,I & 3.25 & $-2.27$ &  3933.7 & Ca\,II & 0.0 & $0.11$ \\
  5270.4 & Fe\,I & 1.6 & $-1.34$ &  3923.5 & Sc\,II & 0.32 & $-2.41$ &  3933.9 & Ti\,I & 2.29 & $-2.32$ \\
  5271.6 & Ti\,I & 2.77 & $-0.87$ &  3924.5 & Ti\,I & 0.02 & $-0.94$ &  3934.2 & Ti\,I & 0.05 & $-2.14$ \\
  5272.0 & Cr\,I & 3.44 & $-0.42$ &  3925.2 & Fe\,I & 3.29 & $-1.4$ &  3935.3 & Fe\,I & 2.85 & $-1.87$ \\
  5273.2 & Fe\,I & 3.29 & $-0.99$ &  3925.6 & Fe\,I & 2.83 & $-1.03$ &  3935.8 & Fe\,I & 2.83 & $-0.88$ \\
  5273.4 & Fe\,I & 2.48 & $-2.16$ &  3925.8 & Sc\,I & 1.99 & $-2.27$ &  3935.9 & Fe\,I & 3.27 & $-2.13$ \\
  5273.4 & Cr\,I & 3.45 & $-0.7$ &  3925.9 & Fe\,I & 2.86 & $-0.94$ &  3936.3 & Ti\,I & 2.48 & $-2.73$ \\
  5274.4 & Ti\,I & 3.09 & $-3.03$ &  3926.0 & Fe\,I & 3.24 & $-0.93$ &  3936.5 & Sc\,I & 2.32 & $-1.6$ \\
  5274.6 & Ti\,I & 2.42 & $-2.97$ &  3926.3 & Ti\,I & 2.58 & $-1.91$ &  3936.8 & Fe\,I & 3.25 & $-1.94$ \\
  5275.2 & Cr\,I & 3.37 & $-0.35$ &  3927.7 & Sc\,I & 1.99 & $-1.36$ &  3937.0 & Ti\,I & 2.29 & $-2.38$ \\
  5275.3 & Cr\,I & 2.88 & $-0.28$ &  3927.8 & Sc\,I & 1.99 & $-1.69$ &  3937.3 & Fe\,I & 2.69 & $-1.46$ \\
  5275.8 & Cr\,I & 2.88 & $-0.05$ &  3927.9 & Fe\,I & 0.11 & $-1.52$ &  3937.6 & Ti\,I & 2.3 & $-2.34$ \\
  5276.0 & Fe\,II & 3.19 & $-2.21$ &  3927.9 & Fe\,I & 2.83 & $-2.27$ &  3937.7 & Fe\,I & 3.02 & $-1.82$ \\
  5276.1 & Cr\,I & 2.88 & $-0.1$ &  3928.1 & Fe\,I & 3.21 & $-0.93$ &  3937.9 & Ti\,I & 2.17 & $-2.28$ \\
  5277.0 & Ti\,I & 3.33 & $-1.67$ &  3928.8 & Ti\,I & 2.31 & $-1.71$ &  3938.0 & Ti\,I & 2.27 & $-2.08$ \\
  5278.2 & Sc\,I & 3.05 & $-1.74$ &  3929.0 & Ti\,I & 2.04 & $-1.5$ &  3938.6 & Sc\,I & 2.3 & $-1.6$ \\
  5278.2 & Ti\,I & 2.34 & $-2.5$ &  3929.0 & Ti\,I & 2.3 & $-1.85$ &  3938.6 & Sc\,I & 2.34 & $-1.72$ \\
  5278.3 & Sc\,I & 3.05 & $-3.05$ &  3929.1 & Fe\,I & 2.76 & $-1.88$ &  3939.7 & Ti\,I & 2.31 & $-2.47$ \\
  5279.7 & Fe\,I & 3.3 & $-3.44$ &  3929.2 & Fe\,I & 3.25 & $-1.34$ &  3940.0 & Sc\,I & 2.61 & $-2.12$ \\
  5280.3 & Cr\,I & 3.36 & $-0.73$ &  3929.9 & Ti\,I & 0.0 & $-1.06$ &  3940.1 & Ti\,I & 2.5 & $-2.96$ \\
  5281.8 & Fe\,I & 3.03 & $-0.83$ &  3930.2 & Sc\,I & 1.99 & $-1.4$ &  3940.9 & Fe\,I & 0.96 & $-2.6$ \\
  5282.4 & Ti\,I & 1.05 & $-1.3$ &  3930.2 & Ti\,I & 1.5 & $-1.23$ &  3941.3 & Fe\,I & 3.27 & $-1.01$ \\
  5283.4 & Ti\,I & 1.87 & $-0.49$ &  3930.3 & Fe\,I & 0.09 & $-1.49$ &  3941.4 & Sc\,I & 2.0 & $-1.5$ \\
  5283.6 & Fe\,I & 3.24 & $-0.43$ &  3930.5 & Fe\,I & 3.21 & $-2.52$ &  3942.4 & Fe\,I & 2.99 & $-2.07$ \\
  5284.1 & Fe\,II & 2.89 & $-3.2$ &  3930.9 & Fe\,I & 2.45 & $-2.86$ &  3942.4 & Fe\,I & 2.85 & $-0.95$ \\
  5284.4 & Ti\,I & 1.04 & $-2.43$ &  3931.1 & Ti\,I & 2.29 & $-2.04$ &  3942.8 & Fe\,I & 3.27 & $-1.31$ \\
  5285.0 & Sc\,I & 2.5 & $-1.42$ &  3931.1 & Fe\,I & 3.27 & $-1.14$ &  3943.3 & Fe\,I & 2.2 & $-2.35$ \\
  5285.6 & Cr\,I & 3.36 & $-1.13$ &  3931.3 & Fe\,I & 3.24 & $-1.9$ &  3943.7 & Sc\,I & 2.0 & $-2.74$ \\
  5285.8 & Sc\,I & 2.5 & $0.38$ &  3931.6 & Ti\,I & 2.32 & $-1.63$ &  3944.3 & Sc\,I & 2.61 & $-2.36$ \\
  \nodata & \nodata & \nodata & \nodata & 3932.0 & Sc\,I & 1.99 & $-1.34$ &  3944.4 & Ti\,I & 2.32 & $-2.83$ \\
  \nodata & \nodata & \nodata & \nodata & 3932.0 & Ti\,II & 1.13 & $-1.65$ &  3944.7 & Fe\,I & 2.85 & $-2.09$ \\
  \nodata & \nodata & \nodata & \nodata & 3932.6 & Sc\,I & 1.85 & $-1.74$ &  3944.9 & Fe\,I & 2.99 & $-1.45$ \\
  \nodata & \nodata & \nodata & \nodata & 3932.6 & Fe\,I & 2.73 & $-1.16$ & \nodata & \nodata & \nodata & \nodata \\
\enddata
 \end{deluxetable}

\end{document}